\documentclass{article}

\usepackage{arxiv}

\usepackage[utf8]{inputenc} 
\usepackage[T1]{fontenc}    
\usepackage{hyperref}       
\usepackage{url}            
\usepackage{booktabs}       
\usepackage{amsfonts, amsmath}       
\usepackage{nicefrac}       
\usepackage{microtype}      
\usepackage{graphicx}
\usepackage{natbib}
\usepackage{array, multirow}
\usepackage{doi}
\usepackage[version=4]{mhchem}

\title{Hybrid Quantum Generative Adversarial Networks To Inverse Design Metasurfaces For Incident Angle-Independent Unidirectional Transmission}

\author{ \href{https://orcid.org/0000-0002-7211-0773}{\includegraphics[scale=0.06]{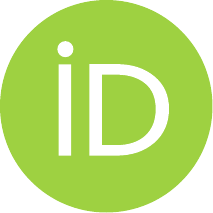}\hspace{1mm}Sreeraj Rajan Warrier} \\
	Department of Physics\\
	Mahindra University\\
	Hyderabad, India 500043 \\
	\texttt{rajan21pphy004@mahindrauniversity.edu.in} \\
	\And
	\href{https://orcid.org/0000-0003-3428-0170}{\includegraphics[scale=0.06]{orcid.pdf}\hspace{1mm}Jayasri Dontabhaktuni} \\
	Department of Physics\\
	Mahindra University\\
	Hyderabad, India 500043 \\
	\texttt{jayasri.d@mahindrauniversity.edu.in} \\
}

\date{}

\renewcommand{\undertitle}{}
\renewcommand{\shorttitle}{\textit{arXiv} Template}

\begin{document}
\maketitle

\begin{abstract}
Optimization of metasurface designs for specific functionality is a challenging problem due to the intricate relation between structural features and electromagnetic responses. Recently, many researchers resolved to inverse design of metasurfaces for efficient design parameters based on methods such as parameter optimization, evolutionary optimization and machine learning. In this paper a hybrid quantum machine learning method which uses quantum encoders to enhance the performance of a classical GAN is applied to implement inverse design of a metasurface. Aiming towards angle-independent unidirectional transmission, this approach combines a Quantum Generative Adversarial Network (Q\-GAN) with a Variational Autoencoder (V\-AE) to optimize metasurface designs. Incident-angle independent uni-directional transmission has potential applications in efficient solar cells, thermal cooling, non-reciprocal devices, etc. However, it is very challenging to achieve such a metasurface design via forward methods and hence very few studies exist till now in this direction. The methodology employed in this work reduces the data requirement for inverse design by 30\% compared to conventional GAN-based methods. More importantly, the developed metasurface designs show high fidelity of 95\% with regard to the targeted far-field radiation patterns. We also provide a material look-up table for feasible substitutes of the obtained material design with real materials and yet maintaining performance accuracy. Further, we embed the inverse-designed metasurfaces into Perovskite solar cell layers to demonstrate the improvement in its performance. We observe that the conversion efficiency of the example perovskite solar cell improves by 95\% and remains independent of incident angle in the range -60$^\circ$ to 60$^\circ$ within the desired frequency range.
\end{abstract}


\section{Introduction}
Designing nanoscale photonic structures that achieve unidirectional transmission of light—robust to variations in the angle of incidence—poses a major challenge in modern optics and quantum nanophotonics. Such functionality demands precise control over both intrinsic material responses and engineered subwavelength architectures, where quantum and classical interference effects play a crucial role in achieving desired asymmetries in wave propagation. Metasurfaces, composed of planar arrays of two-dimensional, subwavelength resonators, have emerged as versatile platforms to control amplitude, phase, and polarization of light at the nanoscale \cite{yu2014flat}. These structures have found broad applicability in areas ranging from beam shaping and holography to sensing, non-reciprocal photonic devices, and thermal radiation control \cite{kildishev2013planar, santos2023directional, Rephaeli2013}.

A key frontier in this domain involves the realization of angle-independent unidirectional transmission. Angle-independent unidirectional transmission has potential applications in designing efficient solar cells, beam steering control, non-reciprocal devices, and thermal cooling \cite{Adam2021, santos2023directional, Chu2024, Rephaeli2013}.
Achieving such behavior necessitates the orchestration of subtle symmetry-breaking mechanisms, resonant mode interference, and optical anisotropy across a broad design space. However, traditional metasurface design techniques, often based on iterative numerical simulations or brute-force parameter sweeps, struggle to identify globally optimal geometries that exhibit this highly constrained and non-trivial functionality \cite{genevet2017recent, dasdemir2023computational}. These methods are computationally expensive, prone to local minima, and limited in their scalability to high-dimensional design tasks \cite{molesky2018inverse}.

To overcome these challenges, inverse design strategies have recently been introduced, aiming to map target optical responses — such as transmission spectral selectivity — to viable metasurface geometries \cite{jiang2019global, Yeung2021global}. In particular, machine learning (ML) approaches, including deep generative models, have shown promise in accelerating and enhancing the inverse design process \cite{liu2018training, So2020}. Such models can learn high-dimensional structure–function relationships and generate diverse candidate solutions without the need for exhaustive simulation. However, classical ML models often require large training datasets and lack inherent mechanisms for exploiting quantum coherence or entanglement that might better capture the underlying quantum-optical phenomena.

In this context, hybrid quantum generative adversarial networks (QGANs) offer a compelling alternative. By leveraging variational quantum circuits embedded within generative frameworks, hybrid QGANs provide a quantum-enhanced mechanism for exploring non-convex design spaces more efficiently and potentially encoding richer structural correlations with fewer parameters. These models have the potential to not only accelerate the inverse design process, but also discover metasurface configurations capable of achieving incident angle-independent unidirectional transmission with high fidelity—an attribute critically important for devices used in beam steering, thermal rectification, and non-reciprocal photonics \cite{sanchez2018inverse, tu2020machine, bingman2020deep, so2019acs}. In this work, we introduce a hybrid QGAN-based inverse design framework specifically tailored to generate metasurfaces that exhibit robust, asymmetric transmission across a wide range of incident angles, opening new directions in the use of quantum-enhanced learning for advanced photonic functionalities.

An effective approach entails the utilization of generative models, including Generative Adversarial Networks (GANs). Generative Adversarial Networks (GANs) signify a notable progression in generative learning models, demonstrating wide-ranging applications in image processing, video analysis, and molecular design \cite{ian2015adversarial}. The models under examination are developed through the training of encoded images that encapsulate material and structural parameters, enabling the creation of metasurface designs that correspond with specified target far-field profiles. Despite these advancements, the complex high-dimensional parameter spaces and intricate design landscapes associated with metasurfaces continue to pose significant challenges for conventional computational methods. This is particularly evident in the issue of mode collapse that is frequently observed in GANs \cite{Kossale2022mode, cerezo2022challenges}. The computational demands of GANs are nearing the thresholds established by Moore's law. Generating images with dimensions of \(512 \times 512\) pixels through the use of GANs necessitates the management of 158 million parameters.   The procedure consists of a training period lasting two days, employing a dataset comprising 14 million instances, and is executed on a configuration of 512 tensor processing units (TPUs) \cite{deng2019unsupervised}.

Quantum algorithms present a unique methodology for the inverse design of metasurfaces, specifically through the application of Quantum Generative Adversarial Networks (QGANs). The networks demonstrate significant potential for enhancing the design process by facilitating efficient navigation and optimization of parameter spaces \cite{preskill2018quantum, lloyd2018quantum}. The combination of quantum computing and GANs enables faster and more efficient convergence to optimal solutions, thus promoting the advancement of accurate and innovative metasurface designs in various technological fields \cite{dallaire2018quantum}. Recent theoretical investigations suggest that quantum generative algorithms may offer an exponential advantage over classical algorithms \cite{mao2017least, wang2018perceptual, sainath2015convolutional}, leading to heightened interest in the theory and experimentation associated with quantum GANs \cite{mao2017least, boyd2004convex, zhang2018self, ioffe2015batch, miyato2018spectral, pande2014numeta}.

Generative Adversarial Networks GANs are fundamentally structured to reproduce the inherent distribution of a specified dataset~\cite{ian2015adversarial}. The simultaneous training of two competing neural networks — generator and discriminator - is conducted to achieve this objective. The generator is designed to create synthetic data samples that closely mimic genuine data, while the discriminator's role is to differentiate between real data and the generated samples. Both networks are compelled to enhance their performance continuously due to this adversarial setup. The generator transforms a latent input distribution \( P_z \) into a generated distribution \( P_g = G(P_z) \), thereby modeling the distribution of real samples represented by \( P_{\text{data}} \). The objective is to render \( P_g \) indistinguishable from \( P_{\text{data}} \); however, achieving this in practice presents significant challenges. A conditional vector in this architecture directs the creation process by incorporating the unidirectionality data. The generator \(G\) and the discriminator \(D\) engage in a minimax optimization process. In this scenario, \(G\) seeks to deceive \(D\) by reducing the probability that \(D\) accurately classifies the generated data, whereas, \(D\) strives to enhance this probability. The objective function for this adversarial training is defined as follows~\cite{Yeung2021global}:

\begin{equation}
\label{eq1}
\min _G \max _D V(D, G) = \mathbb{E}_{\boldsymbol{x} \longrightarrow p_{\text{data}}}[\log D(\boldsymbol{x, \gamma})] + \mathbb{E}_{\boldsymbol{z} \longrightarrow p_z}[\log (1 - D(G(\boldsymbol{z, \gamma})))]
\end{equation}

\begin{itemize}
\item \(\mathbb{E}\) : Expected value
\item \(x\): Real data sample
\item \(z\): Random Noise
\item \(\gamma\): Conditional vector (Radiation Profile)
\item \(D(x)\): Probability that the discriminator accurately identifies real data as authentic
\item \(G(z)\): Generated data
\item \(D(G(z))\): Probability that the discriminator categorizes generated data as authentic
\end{itemize}

The networks undergo iterative training, with each network focused on minimizing a distinct loss function:

\begin{equation}
\label{eq2}
\begin{aligned}
\mathcal{L}_D &= -[y \cdot \log (D(x, \gamma)) + (1 - y) \cdot \log (1 - D(G(z, \gamma)))], \\
\mathcal{L}_G &= -[(1 - y) \cdot \log (D(G(z, \gamma)))]
\end{aligned}
\end{equation}

In this context, \(y\) signifies a binary classification, indicating real data with a value of \(y = 1\) and fake data with a value of \(y = 0\).

In contrast to the original formulation of GANs, which minimizes the expression \(\log[1 - D(G(z, \gamma))]\), our method employs an alternative generator loss function. The traditional formulation is recognized for experiencing vanishing gradients, which obstructs efficient training~\cite{ian2015adversarial, inkawhich_dcgan}.

This research employs hybrid quantum Generative Adversarial Network (QGAN) methodology, referred to, as Latent Style-based Quantum GAN (LaSt-QGAN) \cite{chang2024latent}. The proposed framework integrates two separate components: a pre-trained variational autoencoder \cite{Kingma2013AutoEncodingVB} (VAE) and a patch quantum GAN \cite{Huang2021}. This study outlines an approach that combines the benefits of rapid computational processes with the flexible design characteristics found in image-based deep learning techniques, while also addressing the common issue of mode collapse often seen in GANs. This methodology, based on quantum generative learning applied to imagery, effectively combines the predictability of material properties and structural components with improved computational training and flexible design strategies. Classical GAN has been employed recently to predict the metasurface structure from its absorption spectra \cite{Yeung2021global}. We specifically outline the material and structural properties, which encompass refractive indices and plasma frequencies, as well as their dimensional parameters (such as meta-atom thicknesses) and environmental factors, categorized into three distinct “RGB” channels of color images, as illustrated in prior research \cite{Yeung2021global}.

\section{Methodology} 

\begin{figure}[ht]
\centering
\includegraphics[height=0.5\linewidth, width=\linewidth]{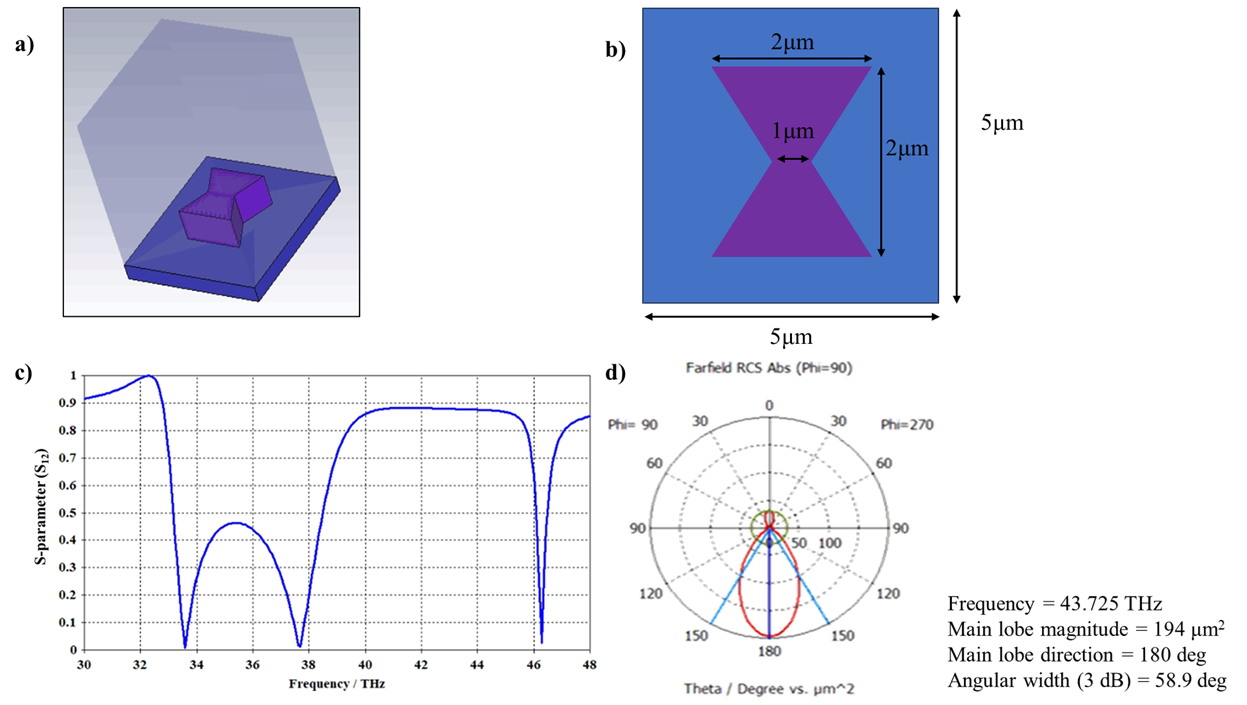}
\caption{Design and characterization of a high-Q metasurface unit cell:
(a) Tellurium bow-tie metasurface cell of $5 \times 5 \times 5.5$ $\mu\text{m}^{3}$  (substrate thickness of 0.5 $\mu\text{m}$ and air thickness of 5 $\mu\text{m}$).
(b) Unit cell of Te bow-tie structure.
(c) Simulated transmission spectrum ($S_{12}$) and 
(d) Corresponding far-field radiation profile. 
}
\label{fig:1}
\end{figure}

This study delves into the inverse design methodology, concentrating on all dielectric unidirectional metasurfaces, as illustrated in Figure~\ref{fig:1}. In this work our focus is exclusively on all-dielectric metasurfaces such as Te, Si and TiO2, within air medium and vacuum. An all-dielectric metasurface is employed over a silica substrate to achieve unidirectionality in visible and infrared (IR) frequencies, where metals typically exhibit significant losses. 
Unidirectional transmission was observed at discrete IR frequencies — 30, 43.725, 56.318, and 60.69 THz, respectively, while in the visible range at various discrete frequencies in the range 334.97 THz to 886.72 THz. For these simulations unit cell of dimensions $5 \times 5 \times 5.5$ $\mu m^3$ is taken as shown in the figure \ref{fig:1}.

\begin{figure}[ht]
\centering
\includegraphics[height=0.3\linewidth, width=\linewidth]{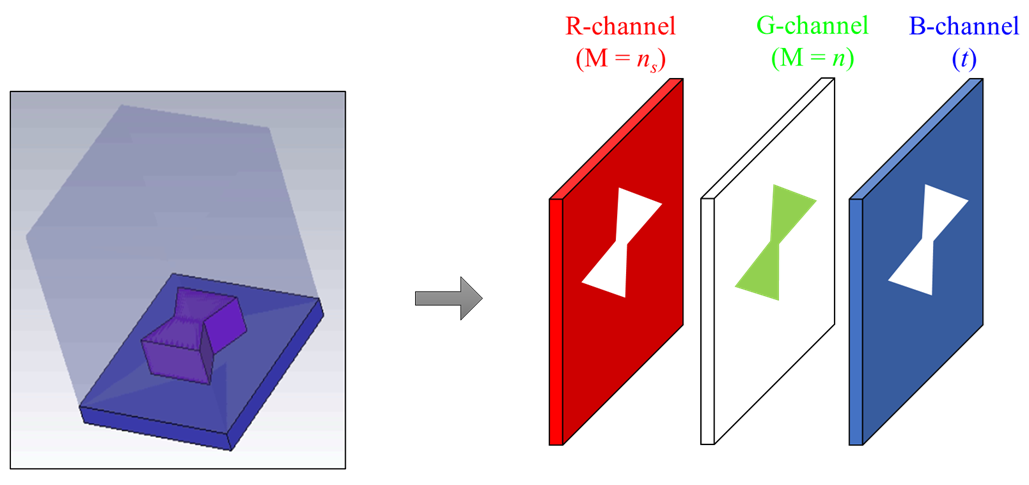}
\caption{Encoding scheme for the unit cell of the metasurface in to RGB channels.}
\label{fig:2}
\end{figure}

The encoding technique depicted in Figure~\ref{fig:2} initiates with the acquisition of planar topologies (G), material characteristics (M), and dielectric layer thicknesses (T) pertinent to both MIM and hybrid dielectric metasurfaces. The parameters under consideration are integrated into three distinct channels of a colored image.The red channel represents the refractive index of the ambient medium - air or vacuum - $(n_s)$, the green channel encodes the refractive index of the resonator material $(n)$, and the blue channel contains information about its thickness $(t)$. This leads to the generation of a red-green-blue channel.

The data set of training pictures for unidirectionality comprises 500 distinct unit cell designs, which are derived from two specific shape templates viz., the bow-tie and the square ring. In the context of unidirectional transmission application, the metasurfaces are restricted to unit cells of size $5 \times 5$ $\mu\text{m}^{2}$ within both air and vacuum environments as shown in the figure \ref{fig:1} a) and b), respectively. The glass substrate thickness is 500 nm and air thickness is 5 $\mu\text{m}$. The constrained availability of qubits presents a significant limitation in our ability to encode the entirety of the values pertaining to the radiation profile within the quantum circuit framework. We select 5 key parameter values, to define the size of the conditional vector $(\gamma)$ for the unidirectionality, aligning with our specific outcome of interest. Transmission spectra given by the scattering parameters $S_{12}$ are calculated and far-field radiation profile is derived at desired frequency using CST Microwave Studio \cite{cst}. From the $S_{12}$ profile as a function of frequency in figure \ref{fig:1} d) we observe that there is a transmission window in the frequency range 38 - 46 THz. Far-field radiation profile in figure \ref{fig:1}d) shows unidirectionality at 43.725 THz with the main lobe magnitude of of 194 $\mu\text{m}^{2}$ directed at 180$^\circ$, and a 3dB angular width of 58.9$^\circ$. The unit cell design is now converted into a three-dimensional pixel representation, characterized as a $64 \times 64 \times 3$ pixel "RGB" image, as depicted in Figure \ref{fig:2}. As a result, a single pixel is associated with a minimum feature size of 79.3 nm for unidirectional metasurfaces, all of which are well within the acceptable limits of fabrication capabilities \cite{Liu2020Compounding, Sunae2019}. Simulations employing the finite element method  were conducted on the generated designs (COMSOL Multiphysics) \cite{comsol} to obtain far-field radar cross section (RCS) radiation profiles for the associated structures.

During the color-encoding phase for the unidirectional metasurfaces, the green channel illustrates the precise refractive indices associated with distinct dielectric resonators, namely (\(n = 2.37\)) for titanium dioxide (\ce{TiO2}), (\(n = 1.449\)) for silica (\ce{SiO2}), (\(n = 4.8\)) for tellurium (\ce{Te}), and (\(n = 9.27\)) for a sample material. Simultaneously, the blue channel consists of height of dielectric structure in the range 100 nm to 1 \(\mu\)m. Additionally, the red channel is utilized to denote the refractive index of the surrounding medium, with values of (\(n_s = 1\)) for vacuum and (\(n_s = 1.0003\)) for air. To preserve the consistency of the "RGB" color system, all embedded values undergo normalization within the range 0 to 255.

\begin{figure}[ht]
\centering
\includegraphics[height=0.5\linewidth, width=0.8\linewidth]{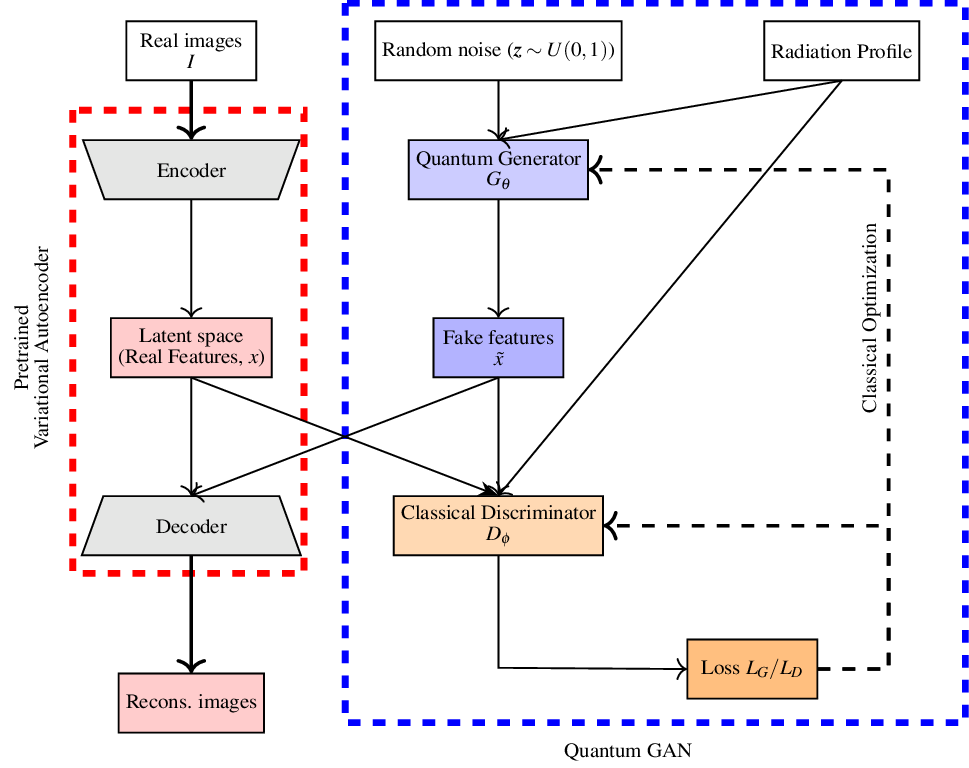}
\caption{Schematic training architecture of a hybrid quantum generative adversarial network (QGAN) with a pretrained variational autoencoder.}
\label{fig:3}
\end{figure}

\begin{figure}[ht]
\centering
\includegraphics[width=\linewidth, height=0.3\linewidth]{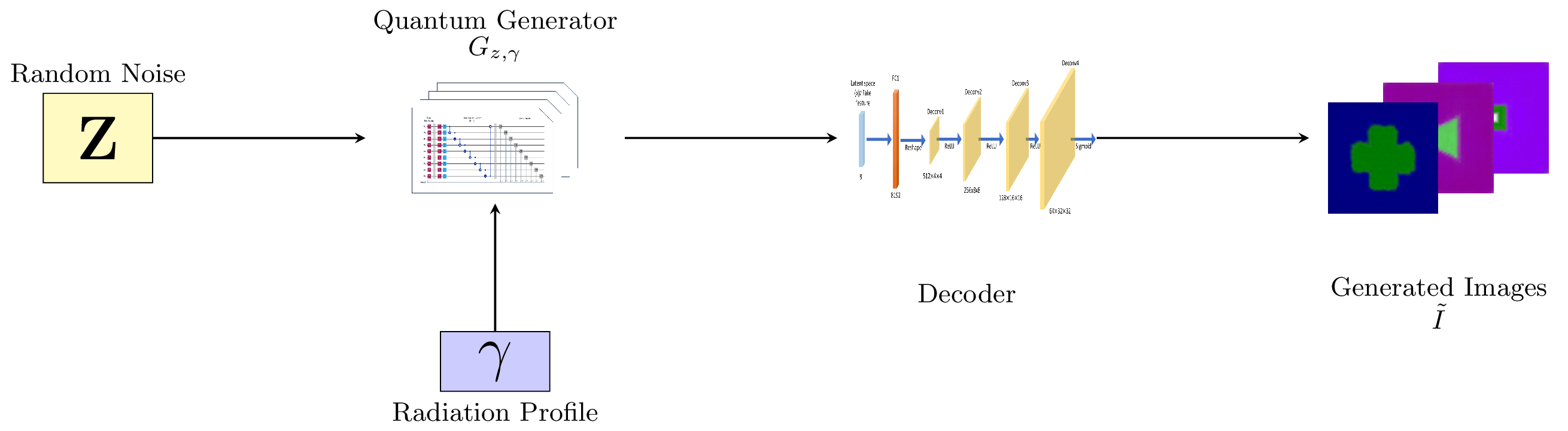}
\caption{Workflow of testing and validation of a hybrid quantum GAN - from latent space sampling to the generation of synthetic metasurface structure images (left to right).}
\label{fig:4}
\end{figure}

The overall workflow for training and inference for LaSt-QGAN are visually summarized in Figures~\ref{fig:3} and~\ref{fig:4}, respectively. Figure~\ref{fig:3} shows the continuous feedback-driven process where the discriminator and generator participate in a non-cooperative game to improve the overall quality of LaSt-QGAN.  During the training process, the input images are transformed using VAE to encode critical information in a compact manner for resource-limited  NISQ quantum computers, instead of learning over the input images pixel by pixel. Figure~\ref{fig:4} highlights the use of decoder of the pretrained VAE to generate metasurface structure images after fully-trained quantum generators. The python script was run on a a GPU server (NVIDIA DGX-1 server with 1 TB RAM, 15TB SSD, 128 CPU cores, and 1 \(\times\) A100 (Amperes) GPUs using Ubuntu 20.04 OS).

In the initial phase, fundamental features denoted as \(x\) are extracted from real images \(I\) within a latent field of predetermined dimensionality, utilizing a standard convolutional variational autoencoder (VAE). The Variational Autoencoder (VAE) is subjected to a pre-training phase that employs the original image dataset, which is defined by dimensions of \(64 \times 64 \times 3\) pixels. This process functions as a reversible method for achieving dimensionality reduction. The characteristics that have been extracted serve as the foundational training collection utilized in the ongoing quantum GAN learning process.

\begin{figure}[ht]
\centering
\includegraphics[height=0.4\linewidth, width=\linewidth]{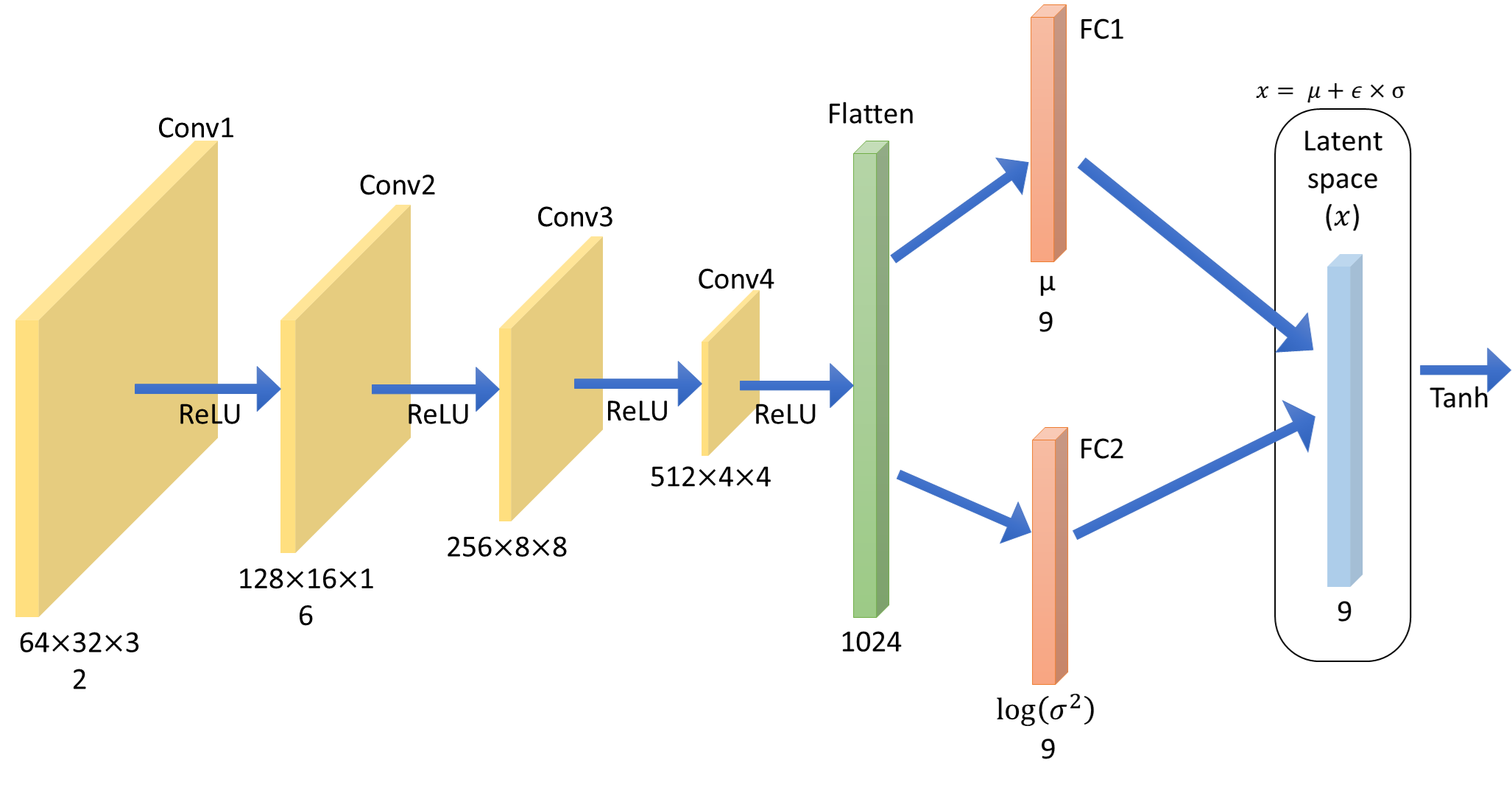}
\includegraphics[height=0.4\linewidth, width=\linewidth]{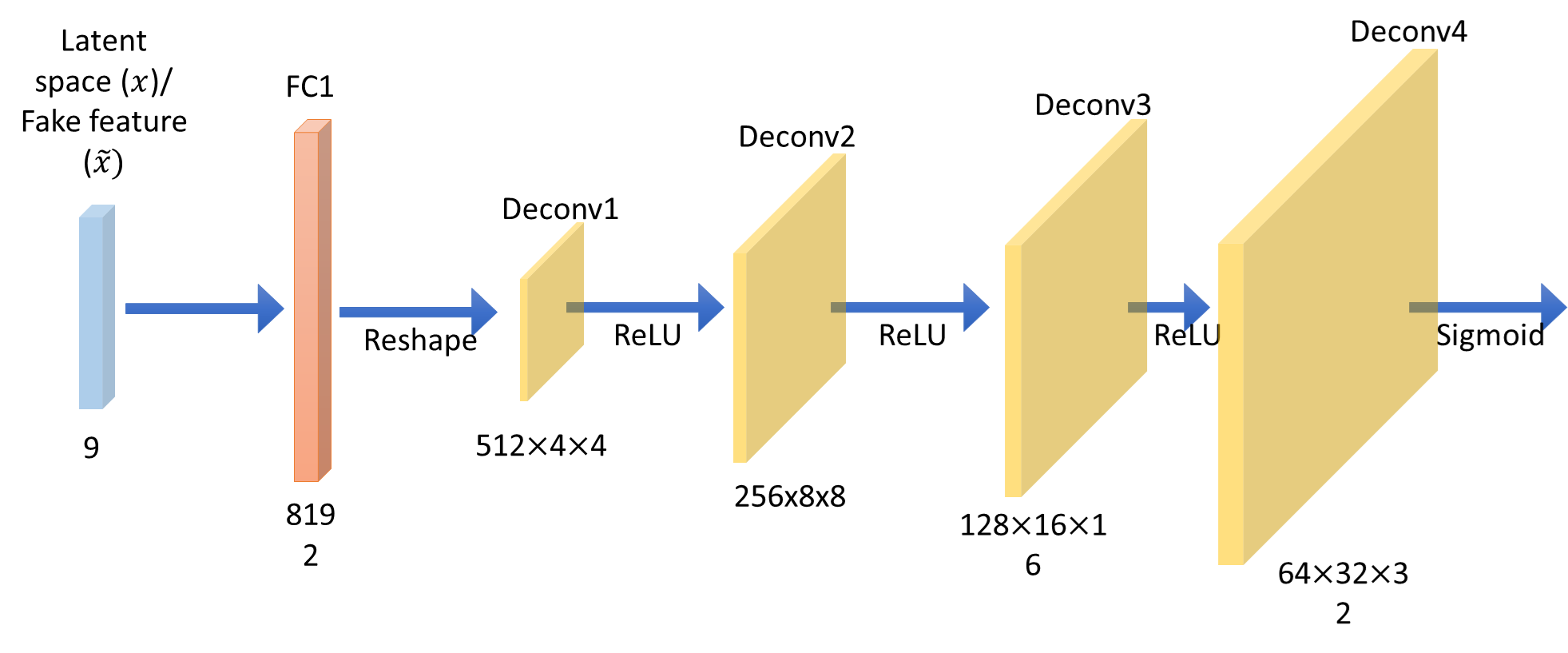}
\caption{Architecture of the $\beta$-VAE and IWAE used in the paper.}
\label{fig:5}
\end{figure}

We compare the results from two pretrained variational autoencoders: $\beta$-VAE and importance weighted autoencoders (IWAE) \cite{higgins2017betavae, burda2015importance}. \(\beta\)-VAE and IWAE are unsupervised neural networks, implemented in the PyTorch Framework, designed for dimensionality reduction and data compression \cite{Adam2019PyTorch}. The key difference between \(\beta\)-VAE and IWAE is in the manner of learning the hidden features of data. \(\beta\)-VAE adds a control parameter, \(\beta\), to balance how well the model reconstructs data versus how well it organizes the hidden features in the latent space, making these features more interpretable. IWAE, on the other hand, uses importance sampling to more accurately capture complex data patterns. In our experiments, we set \(\beta\) to 2 for both the models, and for IWAE, the number of samples is fixed at 5. Figure~\ref{fig:5} illustrates that both VAEs operate by mapping high-dimensional information into a lower-dimensional latent domain and reconstructing data from these latent features (\(x\)) via the decoder, which has a dimension analogous to that of the quantum generator's output.

The loss functions for \(\beta\)-VAE and IWAE are given in eq \ref{eq3} and eq \ref{eq4} below:

\begin{equation}
\label{eq3}
\mathcal{L}_{\text{Beta-VAE}} = \underbrace{\sum_{i=1}^{N} \|x_i - \Tilde{x}_i\|^2}_{\text{Reconstruction Loss}} + \beta \cdot \underbrace{\left(-\frac{1}{2} \sum_{j=1}^{D} \left(1 + \log(\sigma_j^2) - \mu_j^2 - \sigma_j^2 \right)\right)}_{\text{KL Divergence}}
\end{equation}

where,
\begin{itemize}
    \item \( x_i \) is the input data.
    \item \( \Tilde{x}_i \) is the reconstructed data.
    \item \( \mu_j \) and \( \sigma_j^2 \) are the mean and variance from the encoder's output.
    \item \( \beta \) is a weighting factor for the KL divergence term.
\end{itemize}

\begin{equation}
\label{eq4}
\begin{aligned}
\mathcal{L}_{\text{IWAE}} &= -\mathbb{E}_{q_\phi(z|x)} \left[\log \left(\frac{1}{k} \sum_{i=1}^{k} \frac{p_\theta(x, z_i)}{q_\phi(z_i | x)}\right)\right] \\
&= -\mathbb{E}_{q_\phi(z|x)} \left[\log \left(\frac{1}{k} \sum_{i=1}^{k} \exp \left( \underbrace{\log p_\theta(x | z_i)}_{\text{Reconstruction Loss}} + \underbrace{\log p(z_i) - \log q_\phi(z_i | x)}_{\text{KL Divergence}} \right)\right)\right]
\end{aligned}
\end{equation}

where,
\begin{itemize}
    \item \( k \) denotes the total number of importance-weighted samples used in the estimation.
    \item \( p_\theta(x, z) \) refers to the joint probability distribution over the input data and the latent variable.
    \item \( q_\phi(z | x) \) indicates the approximate posterior, estimating the probability of latent variables conditioned on the input.
    \item \( p_\theta(x | z) \) defines the likelihood function, quantifying how well the model reconstructs the input from the latent representation (i.e., reconstruction loss).
    \item \( p(z) \) denotes the prior probability distribution assumed over the latent space.
    \item \( \log p_\theta(x | z_i) \) measures the log-likelihood of the data given a specific latent sample \( z_i \), serving as an indicator of reconstruction quality.
    \item \( \log q_\phi(z_i | x) \) gives the log-probability from the approximate posterior, expressing how likely the latent code \( z_i \) is given the input \( x \).
    \item \( \log p(z_i) \) is the logarithmic form of the prior, representing the assumed probability of the latent code \( z_i \).
\end{itemize}

\sloppy
The first term in the above equation \ref{eq4} gives reconstruction loss and the second term pertains to the relative entropy, given by Kullback-Leibler (KL) Divergence \cite{kullback1951information}.
The quantum Generative Adversarial Network (QGAN) serves as a generative model, where a quantum generator creates synthetic feature representations using random noise in conjunction with a conditional input vector (\(\gamma\)).
In parallel, a classical discriminator differentiates between genuine features (\(x\)) extracted via either a pretrained \(\beta\) Variational Autoencoder (\(\beta\)-VAE) or an Importance Weighted Autoencoder (IWAE), and the synthetic features (\(\Tilde{x}\)) produced by the quantum generator.
The quantum generation process is implemented using the PennyLane simulator developed by Xanadu Technologies \cite{Bergholm2018}.

\begin{figure}[ht]
\centering
\includegraphics[height=0.6\linewidth, width=0.7\linewidth]{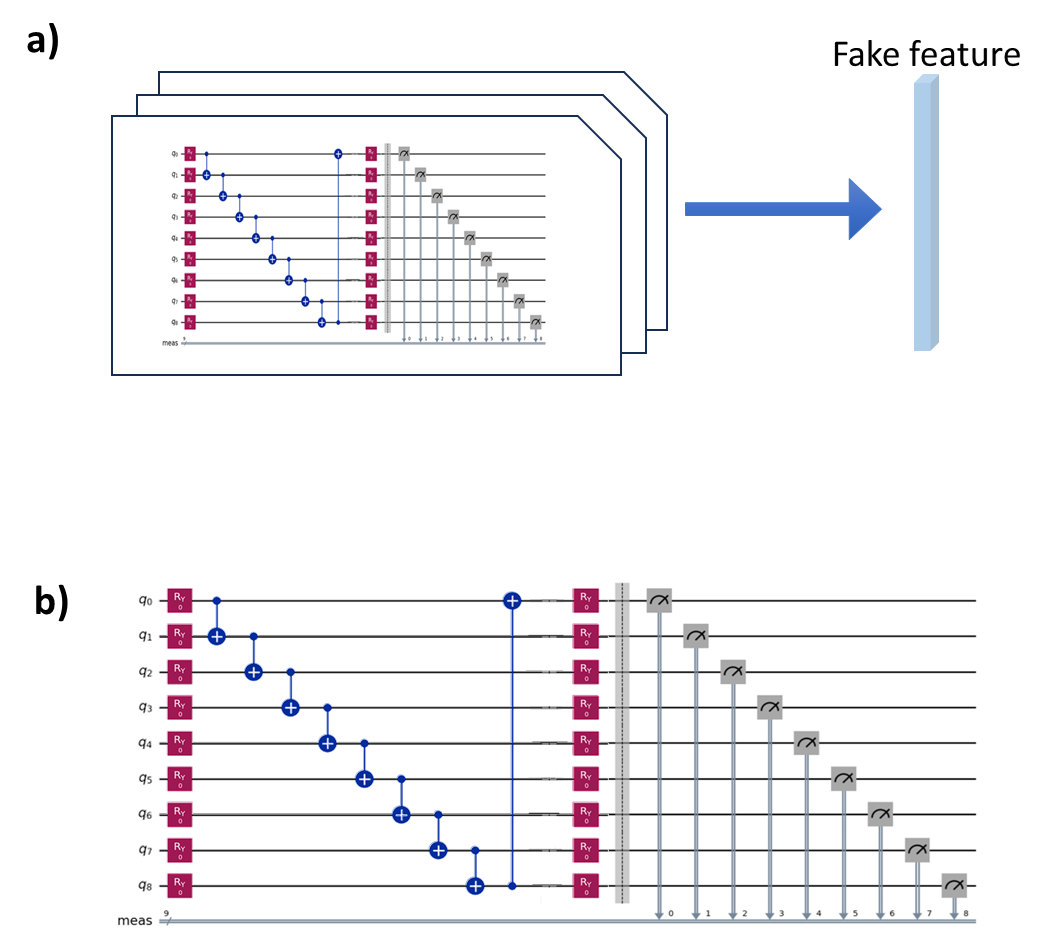}
\caption{Quantum neural network employed for the QGAN model. a) The Patch Quantum Generator connected to a fully-connected layer. b) The quantum circuit employed in each sub-generator.}
\label{fig:6}
\end{figure}

The quantum generator introduced in this study, as shown in Figure~\ref{fig:6}, is designed using a style-based architecture \cite{Bravo-Prieto2022}, where the rotational parameters in the training layers are governed by latent data. This system is configured to handle input data by concatenating the latent noise vector (\(z\)) with conditional information (\(\gamma\)), which is then integrated with the input (\(x\)), facilitating a more dynamic and adaptable learning process. This composite is then directed through a sequence of sub-generators, which are realized as variational quantum circuits \cite{cerezo2021variational, Peruzzo2014}. The iterative refinement of these sub-generators throughout the training process serves to enhance their performance optimization. The circuits exhibit characteristics defined by adjustable parameters, referred to as (\(\theta\)), which are modified during the training process. In the present implementation, the quantum circuit is composed of parameterized layers that incorporate rotation gates denoted as \(\text{RY}\). This configuration is subsequently enhanced by the introduction of circular entanglement through the application of CNOT gates. The comprehensive unitary transformation \( U \) of the circuitry is expressed as \( U = U_\ell(z, \gamma) \cdot \ldots \cdot U_1(z, \gamma) \), wherein each layer \( \ell \) of the circuit incorporates an orthogonal transformation applied to the input \(x\).

\begin{equation}
\label{eq5}
\begin{aligned}
\theta_\ell &= W_\ell \cdot x + b_\ell, \\
\end{aligned}
\end{equation}

where,
\begin{itemize}
    \item \( W_\ell \): The orthogonal weight matrix \cite{kerenidis2021classical} of size \(n_{qubits} \times n_{qubits}\) with $n_{qubits}$ being the number of qubits.
    \item \( b_\ell \): The bias vector of size \(n_{qubits}\).
    \item $x$ : [{$z|\gamma$}] is the concatenation of latent noise ($z$) and conditional vector ($\gamma$).
    \item \( \theta_\ell \): The rotation angles for the quantum gates in the \(\ell\)-th layer of the quantum circuit.
\end{itemize}

In each sub-generator, the input (\(x\)) is integrated into the rotation angles (\(\theta_\ell\)) associated with the quantum gates by the affine conversion specified in equation~\ref{eq5}. The weight matrices are orthogonalized to improve accuracy efficiency and avoid vanishing or inflating gradient problems in quantum neural network topologies, in addition to being uniformly distributed within [-0.01, 0.01] \cite{kerenidis2021classical}. In addition, the bias \(b_\ell\) is selected at random from a uniform distribution spanning [-0.01, 0.01]. The quantum circuit then applies these rotation angles in a sequence of SU2 gates followed by circular entanglement layers, ensuring the proper entanglement of qubits. The entanglement is achieved through the use of CNOT gates, creating a circular pattern of connections between qubits.

During the training process, the model optimizes the set of parameters (\(\Theta = \{W_\ell, b_\ell\}\)) across all layers, adjusting the angles iteratively to improve the generative capacity of the quantum model. The output from the quantum circuit is represented as Pauli-Z expectation values \(\langle \sigma_z \rangle\). This process is repeated across all sub-generators, each contributing to the final output. To ensure consistency, the latent space is also restricted to the interval [-1, 1], matching the output range of \(\langle \sigma_z \rangle\) from the quantum generator. This architectural framework facilitates the representation of intricate, high-dimensional data, rendering it appropriate for a diverse array of tasks within the domain of quantum machine learning.

The classical discriminator employed in this research is a fully connected neural network, specifically engineered to differentiate between the latent space ($x$) and the synthetic feature representation (\(\Tilde{x}\)) within the framework of the Quantum Generative Adversarial Network (QGAN). It takes input features, along with corresponding conditional vector (\(\gamma\)), and processes them through a sequence of layers. Initially, the conditional vector data undergoes a linear transformation to match the dimensionality of the input features. The classical discriminator consists of three fully connected layers, with the first layer having 128 nodes, followed by a second layer with 64 nodes. Each of these layers is followed by a Leaky ReLU activation function \cite{Jin2020ReLu} with a negative slope of 0.2. The final layer comprises a solitary node, succeeded by a Sigmoid activation function \cite{Cybenko1989}, which yields a probability score that signifies the authenticity of the input data, determining whether it is authentic or fake. This feedback mechanism is crucial for guiding the quantum generator towards producing more realistic features.

\section{Results and Discussion}

\begin{figure}[ht]
\centering
\includegraphics[height=0.6\textwidth,width=\textwidth]{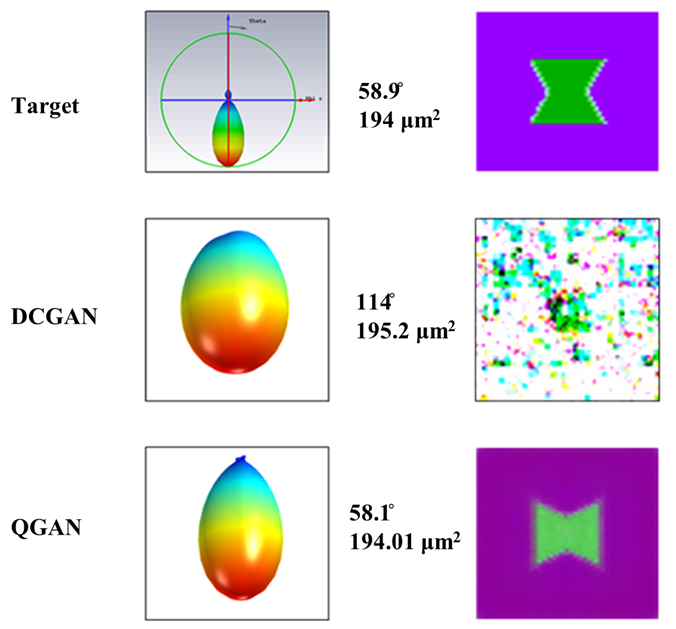}
\caption{Comparison of the far-field profiles and the unit cell designs of the target, DCGAN model, and LaSt-QGAN model, respectively.}
\label{fig:7}
\end{figure}

\begin{figure}[ht]
\centering
\includegraphics[width=\textwidth]{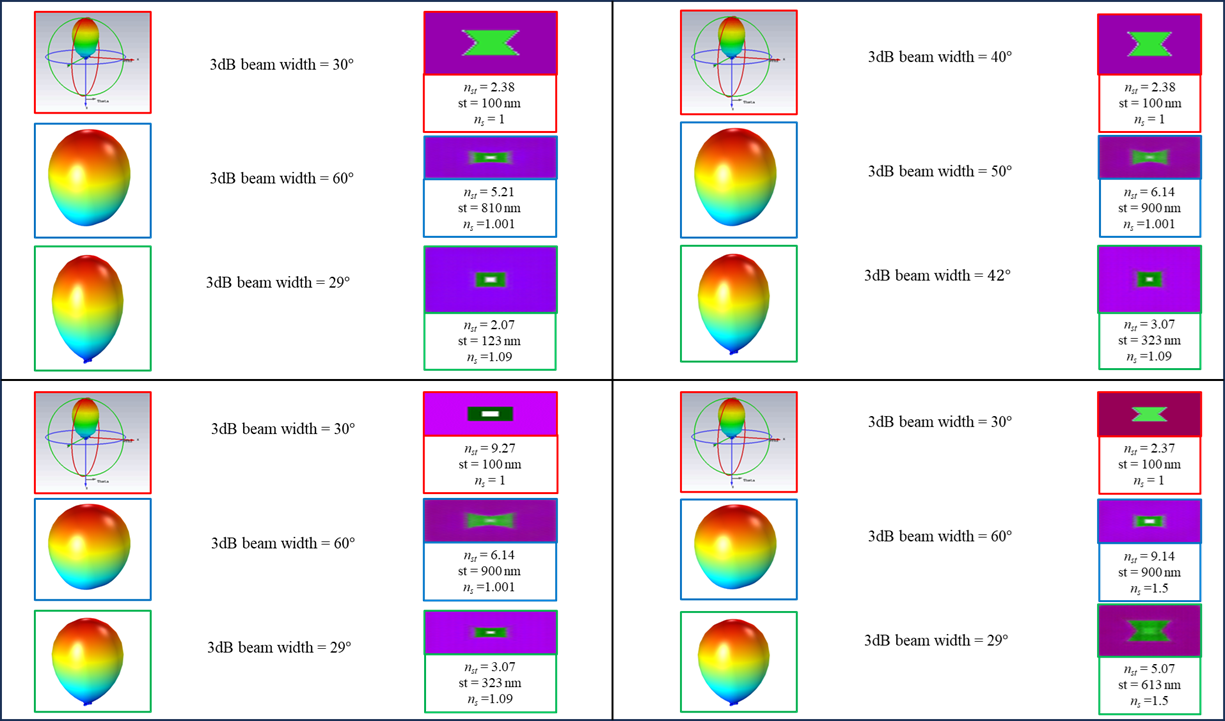}
\caption{Simulated 3D far-field profiles from 4 sets of La-St QGAN designs using pretrained \(\beta\)-VAE and IWAE  (blue and green outlined boxes), respectively, compared to the target far-field profiles (in red).}
\label{fig:8}
\end{figure}

Using the obtained values for refractive index (\(n\)), new materials were defined within the EM simulations for the QGAN-based metasurface design process. It is essential to recognize that conventional manufacturing methodologies may prove inadequate for the novel materials produced via this approach. The material definition framework allows the model to explore more design possibilities by predicting a wider range of material properties, which might otherwise be overlooked if simplified into fixed categories.
First we compare the far-field profile results from the LaSt-QGAN model to the target and DCGAN model as shown in Figure~\ref{fig:7}. The values given in Figure~\ref{fig:7} are the angular width at 3 dB and the main lobe magnitide value. The runtime to train the LaSt-QGAN model is 2.5 hours. In order to assess the efficacy of the trained La-St QGAN and the employed image processing methodology, we introduced a collection of far-field radiation profiles in conjunction with arbitrary latent vectors. The subsequent analysis of the resultant designs, as illustrated in Figure~\ref{fig:4}, pertains to the aspect of unidirectionality, respectively. In accordance with the outlined methodology, Figure~\ref{fig:8} illustrate a sequence of experiments utilizing inputs derived from the validation dataset, which constitutes 10\% of the training data. The red box denote inputs that have been selected at random for unidirectional metasurfaces. In contrast, the blue and green box illustrate the simulated far-field radiation profiles corresponding to designs generated by La-St QGAN, utilizing pretrained \(\beta\)-VAE and IWAE methodologies, respectively. Images of the corresponding unit cells are displayed adjacent to each plot.

\begin{figure}[ht]
\centering
\includegraphics[width=\textwidth]{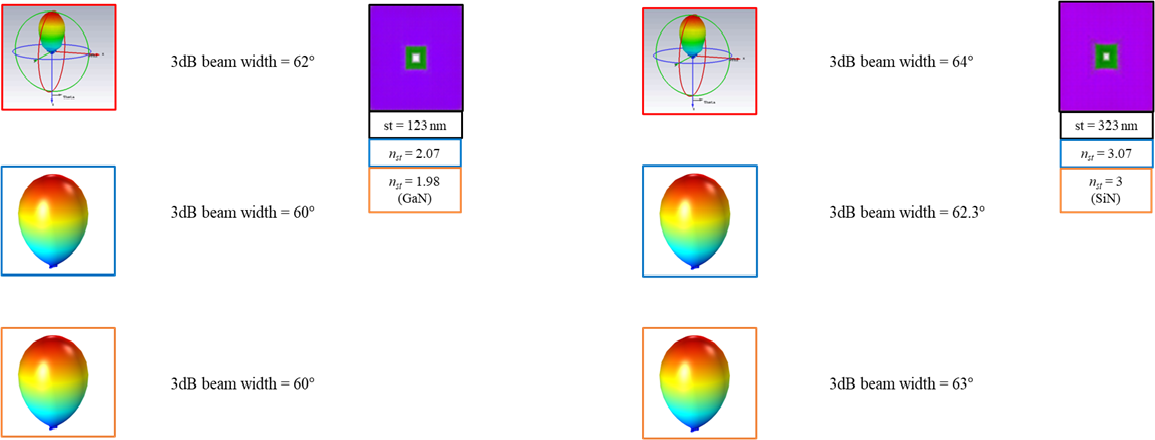}
\caption{The target far-field radar cross-section (RCS) profile (red box) is compared with simulated results using approximate materials (orange box) and QGAN-predicted materials (blue box).}
\label{fig:9}
\end{figure}

To improve the practical manufacturability of the metasurface designs, we integrated a look-up table by substituting the refractive index values produced by the QGAN with those of the nearest matching materials obtained from a publicly accessible database \cite{polyanskiy2024refractiveindex}. Substitutions comprising of materials such as gallium nitride (\ce{GaN} , \(n\) = 1.99) and silicon carbide (\ce{SiC}, \(n\) = 3) are adopted for unidirectional metasurfaces as shown in Figure~\ref{fig:9}. Re-simulating these designs demonstrated that the approximated material properties maintain about 90\% precision, when compared to the original predictions generated by the GAN. This finding demonstrates a viable alternative within the inverse design framework to achieve more accessible and manufacturable designs without compromising performance accuracy. Furthermore, it is notable that some materials identified through this process, such as \ce{GaN} and \ce{SiC}, were not included in the original training dataset. This underscores the capability of the QGAN to predict new material parameters — such as refractive index, that identifies suitable alternative materials meeting the target specifications, even if they were absent from the training data.

The QGAN-based approach underscores its robustness by offering a level of generalization and design flexibility that surpasses traditional machine learning techniques, which are constrained to predicting materials strictly within the boundaries of the training dataset. In contrast, our methodology enables the exploration of material alternatives that extend beyond these predefined limits. Although the examples provided demonstrate the QGAN's ability to generate predictions aligned with known materials, it is recognized that the model might also estimate characteristics beyond the established limits of standard materials. Nevertheless, it is expected that the precision of these predictions will increase as material databases grow and the variety of available materials broadens. From Table~\ref{tab:1}, we observe that both DCGAN and QGAN take approximately similar times (2 hrs and 2.5 hrs, respectively) to train a reasonably smaller sample size of 500 for 500 epoch cycles. QGAN, however, outperforms DCGAN in accuracy, highlighting the potential of quantum generative models for tasks requiring high fidelity, especially in low-data regimes. We would like to note that while 500 sample size is sufficient for QGAN to efficiently generate the metasurface design, DCGAN on the other hand, requires 20000 samples for training the data to generate a good quality metasurface design \cite{Yeung2021global}.

\begin{table}[htbp]
    \centering
    \begin{tabular}{c|c|c}
         \hline
         \textbf{Metric} & \textbf{QGAN} & \textbf{DCGAN} \\
         \hline
         Runtime (hrs)     & 2.5 & 2.01 \\
         MSE               & $10^{-4}$ & $10^{-1}$ \\
         \hline
    \end{tabular}
    \caption{Performance metrics of the QGAN and DCGAN model, including runtime, mean squared error (MSE).}
    \label{tab:1}
\end{table}

In addition to demonstrating the unit cell design for incident angle-independent unidirectional transmission, we also show how the QGAN generated metasurface can enhance the efficiency of a solar cell. To demonstrate this a perovskite solar cell with RGO contact is taken with the inverse-designed metasurface embedded in the cell for enhanced absorption in the solar cell. It was shown earlier that such a metasurface with incident-angle independent uni-directional transmission enhances the solar cell conversion efficiency \cite{iqbal2024metasurface}. In the current work, COMSOL Multiphysics is used for end-to-end solar cell simulations \cite{Zandi2020}. We introduce the inverse-designed metasurface over a glass (silica) substrate and sandwich between FTO and glass (silica) substrates, while rest of the layers remain the same. We then simulate the solar cell of dimensions \(250 \times 250\) \(\text{nm}^2\) in the frequency range 300 to 900 nm \cite{Zandi2020}. Unit cell of the metasurface is designed to have same 2D dimensions as the solar cell. Incident light through the metasurface undergoes unidirectional transmission independent of angle of incidence in the frequency range considered. Data from AM1.5G is used for incident solar power for the plane wave incidence in the range 300–900 nm \cite{nrel_am1.5g}.

\begin{figure}[ht]
\centering
\includegraphics[height=0.65\textwidth, width=0.85\linewidth]{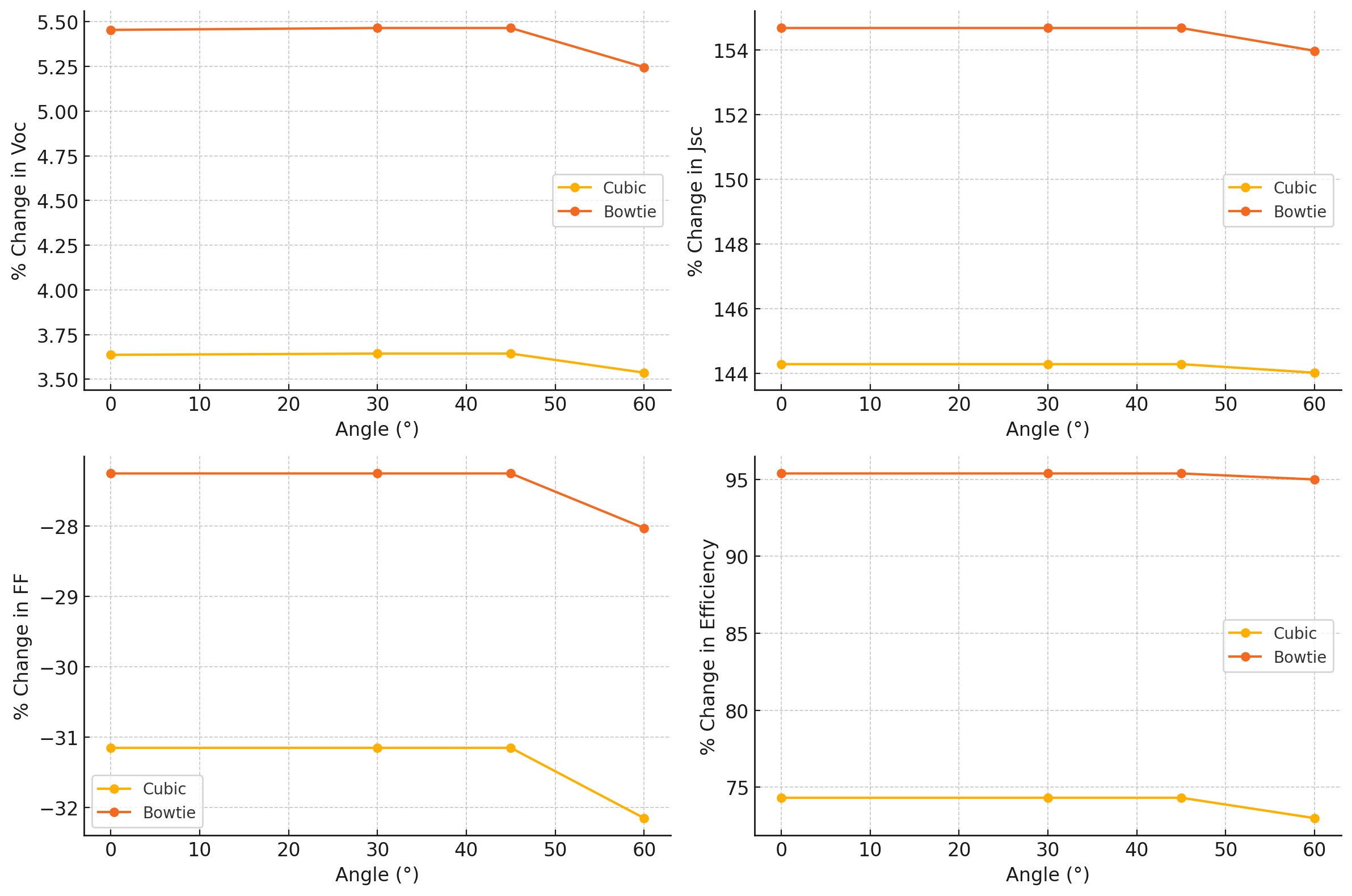}
\caption{Performance metrics \(V_{oc}\), \(J_{sc}\), FF, and efficiency as a function of the angle of incidence for cubic and bowtie metasurface designs.}
\label{fig:10}
\end{figure}

The parameters of interest include the open-circuit voltage \(V_{oc}\), short-circuit current density \(J_{sc}\), fill factor \(FF\), and overall efficiency \(\eta\). Figure~\ref{fig:10} illustrates the angular dependence of each metric, comparing the baseline (without metasurface) solar cell with two metasurface designs: Cubic and Bowtie. The metasurfaces result in notable increase in Jsc and efficiency, as demonstrated. A very minor reduction in performance is noted at oblique angles (normal incidence corresponds to angle $0^\circ$), showcasing an incident-angle independence within a wide range of angles $-60^\circ$ till $60^\circ$ .Figure~\ref{fig:10} shows that bowtie metasurfaces demonstrate superior performance compared to cubic metasurfaces in terms of Voc, Jsc, and overall efficiency enhancements across all incidence angles, achieving efficiency improvements of up to 95\%. Both designs show a reduction in FF after $60^\circ$, confirming the angular robustness of solar cells enhanced by metasurfaces.

Table~\ref{tab:2} lists the main performance indicators for baseline, cubic, and bowtie metasurface solar cells at various angles of incidence. These include open-circuit voltage (\(V_{oc}\)), short-circuit current density (\(J_{sc}\)), fill factor (\(FF\)), and efficiency (\(\eta\)). According to the data, bowtie metasurfaces consistently exhibit superior efficiency at all angles, with an enhancement of up to 95\% . Cubic metasurfaces also show notable increase, especially in \(J_{sc}\). Even though both metasurface designs have lower \(FF\), the significant gains in \(J_{sc}\) and \(V_{oc}\) offset the losses and improve overall performance.  The metasurfaces are also angularly resistant, showing efficiency gains even at oblique angles, which make them attractive options for real-world solar applications. To demonstrate the performance improvement of perovskite solar cell we used inverse-designed metasurfaces obtained by employing quantum GAN methods in the current work. However, it should be noted that this can be extended to other photovoltaic cell designs such as based on Si and other heterogenous designs as well.

\begin{table}[ht]
    \centering
    \renewcommand{\arraystretch}{} 
    \caption{Performance comparison of \(V_{oc}\), \(J_{sc}\), FF, and Efficiency for Baseline, Cubic, and Bowtie metasurfaces at different angles of incidence.}
    \resizebox{0.85\textwidth}{!}{ 
    \begin{tabular}{c|ccc|ccc}
        \hline
        \multirow{2}{*}{{Angle (°)}} & \multicolumn{3}{c|}{{Voc (V)}} & \multicolumn{3}{c}{{Jsc (mA/cm\(^2\))}} \\
        \cline{2-7}
        & {Baseline} & {Cubic} & {Bowtie} & {Baseline} & {Cubic} & {Bowtie} \\
        \hline
        0  & 1.1 & 1.14 & 1.16 & 17.472 & 42.683 & 44.496 \\
        30 & 1.043 & 1.081 & 1.1 & 14.48657 & 35.38979 & 36.893 \\
        45 & 1.043 & 1.081 & 1.1 & 12.92848 & 31.58346 & 32.925 \\
        60 & 1.07368 & 1.11165 & 1.13 & 11.81216 & 28.82515 & 29.999 \\
        \hline
        \multirow{2}{*}{{Angle (°)}} & \multicolumn{3}{c|}{{FF}} & \multicolumn{3}{c}{{\(\eta\) (\%)}} \\
        \cline{2-7}
        & {Baseline} & {Cubic} & {Bowtie} & {Baseline} & {Cubic} & {Bowtie} \\
        \hline
        0  & 0.486857 & 0.335192 & 0.354197 & 9.357 & 16.31 & 18.282 \\
        30 & 0.437741 & 0.301377 & 0.318464 & 6.614696 & 11.52994 & 12.924 \\
        45 & 0.431518 & 0.297092 & 0.313937 & 5.819335 & 10.14357 & 11.37 \\
        60 & 0.437686 & 0.296962 & 0.315025 & 5.476583 & 9.473563 & 10.679 \\
        \hline
    \end{tabular}
    } 
    
    \label{tab:2}
\end{table}

\section{Conclusions}

This study employs Latent Style-based Quantum GAN (LaSt-QGAN) as a hybrid quantum-classical framework for the inverse design of metasurfaces. The model enhances metasurface structures to achieve unidirectionality and simultaneously lowers computational expenses by incorporating a Quantum Generative Adversarial Network (QGAN) alongside a Variational Autoencoder (VAE). LaSt-QGAN achieves 95\% fidelity in relation to the target far-field radiation patterns. Furthermore, it minimizes data requirements by 30\% in comparison to conventional GAN-based approaches highlighting the potential of quantum generative models for tasks requiring high fidelity in data generation, especially in low-data regimes. The experimental validation of metasurface-enhanced perovskite solar cells demonstrates significant performance improvements, particularly in cubic and bowtie geometries, which show enhanced unidirectional properties. A material look-up table facilitates practical manufacturability by substituting predicted materials with real-world alternatives, while ensuring accuracy is preserved.

The integration of the generated meta-structures into a perovskite solar cell results in notable improvements in performance, in conjunction with the design of the metasurface. The integration of metasurfaces, when compared to a baseline solar cell, results in improvements in overall power conversion efficiency (\(\eta\)), short-circuit current density (\(J_{sc}\)), and open-circuit voltage (\(V_{oc}\)).  The cubic metasurface achieves an efficiency of 16.31\%, whereas the bowtie metasurface enhances this to 18.28\%. The observed enhancements demonstrate that the inverse-designed metasurfaces have the potential to enhance the efficiency of solar energy harvesting.

Furthermore, the angle-dependent analysis of metasurface-enhanced solar cells demonstrates that performance improvements are maintained across various incident angles, with efficiency enhancements surpassing 70\% at oblique incidences of even 45° and 60° when compared to the baseline. The durability of these designs demonstrates their suitability for real-world applications in scenarios where the intensity of incoming sunlight fluctuates throughout the day. The observed trends demonstrate that cubic and bowtie metastructures significantly decrease angular sensitivity while maintaining high power conversion efficiency, underscoring their suitability for practical photovoltaic systems.

This article demonstrates the accelerated inverse design methodologies for metasurfaces achievable via quantum-enhanced machine learning methods, while maintaining a focus on practical fabricability. The observed enhancement in solar cell efficiency is a result of presence of inverse-designed metasurface achieving incident angle-independent unidirectional transmission and underscores the possible applications in energy harvesting, sensing, imaging, and telecommunications. This method mitigates computational inefficiencies and mode collapse in Generative Adversarial Networks (GANs) while providing improved flexibility for investigating high-dimensional design spaces. The capability of quantum machine learning to revolutionize photonics and materials science is underscored by the efficiency of this approach in enhancing metasurfaces. Future research in this context could focus on the investigation of adaptive metasurface designs that can dynamically optimize the performance of solar cells in response to varying illumination conditions, which encompasses changes in both spectral, intensity and angular distributions of sunlight. Furthermore, substantial progress in the advancement of next-generation optical and energy-harvesting technologies may be realized by expanding this methodology to include multi-functional metasurfaces and a broader spectrum of photonic applications.

\section*{Acknowledgments}
This research is supported and financed by Mahindra University.

\bibliographystyle{unsrt}
\bibliography{references}

\end{document}